\documentclass{aa}
\input{psfig.sty}
\usepackage{lscape}
\usepackage{graphicx,times}
\usepackage{float}
\usepackage{longtable}

\begin{document}
\authorrunning{K. C. Steenbrugge et al.}
\titlerunning{Heavily absorbed spectrum on IC 4329A}
\title{XMM-Newton observations of the heavily absorbed Seyfert 1 galaxy IC 4329A}
\subtitle{}

\author{K.C. Steenbrugge\inst{1}, J.S. Kaastra\inst{1}, M. Sako\inst{2}, G. Branduardi-Raymont\inst{3}, E. Behar\inst{4}, F. B. S. Paerels\inst{5}, A. J. Blustin\inst{3} and S. M. Kahn\inst{2}}
\offprints{K.C. Steenbrugge}
\mail{K.C.Steenbrugge@sron.nl}
\institute{ SRON National Institute for Space Research, Sorbonnelaan 2, 3584 CA Utrecht, The Netherlands
\and Kavli Institute for Astrophysics and Cosmology, Stanford University, 382 Via Pueblo Mall, Stanford, CA 94305-4060, USA
\and Mullard Space Science Laboratory, University College London, Holmbury St Mary, Dorking Surrey RH5 6NT, UK
\and Physics Department, Technion 32000, Haifa, Israel
\and Columbia Astrophysics Laboratory, Columbia University, 550 West 120th street, New York, NY 10027, USA}
\date{Received  / Accepted  }

\abstract{We detect seven distinct absorbing systems in the high-resolution X-ray spectrum of the Seyfert 1 galaxy IC 4329A, taken with XMM-{\it Newton}. Firstly we detect absorption due to cold gas in our own Galaxy and warm gas in the Galactic halo or the Local Group. This local warm gas is only detected through \ion{O}{vii} absorption, from which we deduce a temperature between 0.03 and 0.2 keV. In IC~4329A we detect absorption from the host galaxy as well as from a warm absorber, close to the nucleus, which has 4 components. The absorption from the host galaxy is well modeled by neutral material. The warm absorber detected in IC~4329A is photoionized and has an ionization range between log $\xi$ = $-$1.37 and log $\xi$ = 2.7. A broad excess is measured at the \ion{O}{viii} Ly$\alpha$ and \ion{N}{vii} Ly$\alpha$ emission lines, which can be modeled by either disklines or multiple Gaussians. From the lightcurve we find that the source changed luminosity by about 20 \% over the 140 ks observation, while the spectral shape, i.e. the softness ratio did not vary. In the EPIC spectra a narrow Fe K$\alpha$ and \ion{Fe}{xxvi} Ly$\alpha$ emission line are detected. The narrowness of the Fe K$\alpha$ line and the fact that there is no evidence for flux variability between different observations leads us to conclude that the Fe K$\alpha$ line is formed at a large distance from the central black hole.
}

\maketitle

\section{Introduction\label{sect:intro}}
IC~4329A is one of the X-ray brightest Seyfert 1 galaxies in the sky. The spiral galaxy is seen almost edge on and is the closest companion to the elliptical galaxy IC~4329, the central galaxy in a cluster. IC~4329A has rather extreme properties,  such as the full width at zero intensity (FW0I) of the H$\alpha$ line which equals 13,000 km~s$^{-1}$ (Disney, \cite{disney}). On the basis of the reddening observed in the optical, $A_v$ = $2.5 - 4.8$ magnitudes, Wilson \& Penston (\cite{wilson}) have classified IC~4329A as the nearest quasar, with an absolute magnitude between $-$23.0 and $-$25.3. In the optical spectra prominent \ion{Na}{i} D, \ion{Ca}{ii} and weaker \ion{Mg}{i} absorption lines are observed, indicating the presence of dust, consistent with the dust lane observed in the equatorial plane of the galaxy (Wilson \& Penston \cite{wilson}). The optical spectrum shows that the absorber is neutral or very lowly ionized. The observed line emission is due to lowly ionized material with strong [\ion{O}{iii}] (Wilson \& Penston \cite{wilson}). Crenshaw \& Kraemer (\cite{crenshaw}) classified IC~4329A as a dusty luke-warm absorber on the basis of similarities with the known absorbers of this type in NGC~3227 and Ark~564. These similarities are the high inclination angle and the reddening determined between the far UV and optical band. Crenshaw \& Kraemer (\cite{crenshaw}) predict that only lowly ionized absorption should be detected, namely the dust that causes the reddening observed in the optical and the UV band. 
\par
The redshift of IC~4329A is 0.01605 $\pm$ 0.00005 (Willmer et al. \cite{willmer}). However, Wilson \& Penston (\cite{wilson}) note that the optical lines are double peaked, with the second peak at a mean redshift of 0.0224 $\pm$ 0.0001. Throughout our analysis we fixed the redshift to $z$ = 0.01605, as we did not detect any evidence for a component at z = 0.0224. All spectra were corrected for a Galactic hydrogen absorption column density $N_{\rm H}$ = 4.55 $\times$ 10$^{24}$ m$^{-2}$ (Elvis, Lockman \& Wilkes \cite{elvis89}). 
\par
Earlier X-ray spectra contained a reflection component (Miyoshi et al. \cite{miyoshi}) and a moderately broadened Fe K$\alpha$ emission line (Done et al. \cite{done}). In an earlier 12 ks XMM-{\it Newton} observation Gondoin et al. (\cite{gondoin}) detected only a narrow Fe K$\alpha$ line, with a width $\sigma$ = 0.01 $\pm$ 0.05 keV. The earlier {\it Chandra} HEG spectrum shows a complex double line at the Fe K$\alpha$ energy, which can be fit either by two Gaussians, two disklines, or a single diskline (McKernan \& Yaqoob \cite{mckernan}). Gondoin et al. (\cite{gondoin}) fitted the soft X-ray part of the spectrum with a set of absorption edges of \ion{O}{i}, \ion{O}{vi}, \ion{O}{viii} and \ion{N}{vii}. They noted the lack of an \ion{O}{vii} absorption edge and concluded that the warm absorber must have at least 2 phases.
\par
In this paper we analyze a deep 140 ks XMM-{\it Newton} observation of IC~4329A. In Sect. 2 we describe the observation and data reduction. The variability is analyzed in Sect. 3. In Sect. 4 the spectral analysis is detailed, while Sect. 5 details the analysis of the absorbers present in the spectrum. In Sect. 6 we discuss our results and the conclusions are summarized in Sect. 7. Power density spectra analysis as well as possible variability in the Fe K$\alpha$ line will be described by Markowitz et al. (\cite{markowitz}) once the complete RXTE long term monitoring campaign is completed.

\section{Observation and data reduction}
IC~4329A was observed on August 6 and 7, 2003, for a total exposure time of 136 ks. Throughout the observation there were periods of high background, which were removed for the EPIC data using the Good Time Interval (GTI) produced with the online Processing Pipeline Subsystem PPS data files. For the RGS spectra we use the whole exposure time, to maximize the statistical quality of the data. For the data reduction we used the PPS data products, downloaded from the XMM-{\it Newton} website. The data were reduced with the SAS software version 5.4.1. Table~\ref{tab:tab1} lists the instrumental set-up and the net exposure time used in the further spectral analysis. For the time variability analysis we used the full 140 ks pn observation. The pn, MOS 1 and MOS 2 spectra were fitted between 0.4 $-$ 12 keV, the first order RGS spectra were originally fitted between 7 $-$ 38 \AA, the second order spectra between 7 $-$ 19 \AA. Due to the low count rate, we limited our analysis to the first order spectra of RGS, and the 7 to 26 \AA~wavelength range. RGS 1 and 2 were averaged in the presented plots for clarity, but fit as separate spectra. For both RGS instruments the count rate ranged between 30 and 250 counts per bin for the part of the spectrum up to 26 \AA, allowing the use of $\chi^2$ statistics. The RGS and MOS spectra were binned by a factor of 3, for the RGS this corresponds to about half the FWHM resolution of 0.07 \AA~or a bin size of 0.03 \AA. The MOS cameras have a FWHM resolution of 50 eV at 0.4 keV and 180 eV at 10 keV. The pn data were binned by a factor of 5, and have a FWHM resolution ranging between 90 eV and 185 eV at 1 keV and 10 keV respectively. All the spectra were analyzed using the SPEX software (Kaastra et al. \cite{kaastrasp}). All quoted errors are rms errors, i.e. $\Delta \chi^2$ = 2.

\begin{table}
\begin{center}
\caption{Instrumental set-up and exposure time used for the spectral analysis.}
\label{tab:tab1}
\begin{tabular}{|lll|} \hline
Instrument  & exposure (s)  & observational mode \\\hline
RGS 1 and 2 & 136017        & spectroscopy \\
pn          & 44363         & Small window with thin filter \\
MOS 1       & 97857         & PrimePartialW3 with medium filter \\
MOS 2       & 94709         & Small window with thin filter \\\hline
\end{tabular}
\end{center}
\end{table}

\section{Time variability}
\begin{figure}
 \resizebox{\hsize}{!}{\includegraphics[angle=-90]{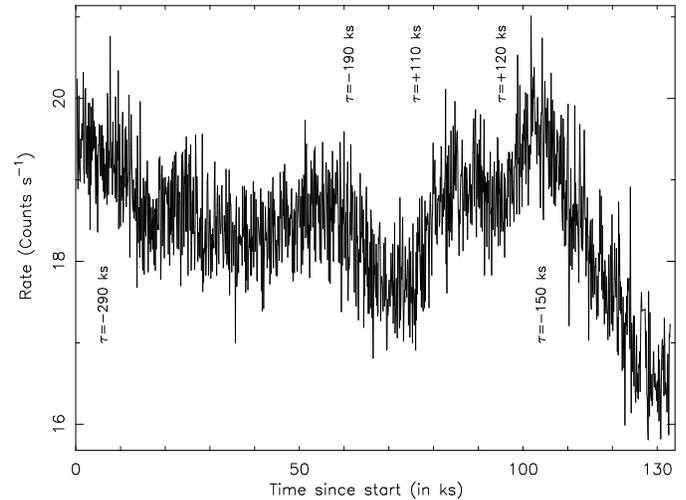}}
\caption{The background subtracted lightcurve for the pn instrument, using the complete 140 ks observation.}
\label{fig:pn_light}
\end{figure}

\begin{figure}
 \resizebox{\hsize}{!}{\includegraphics[angle=-90]{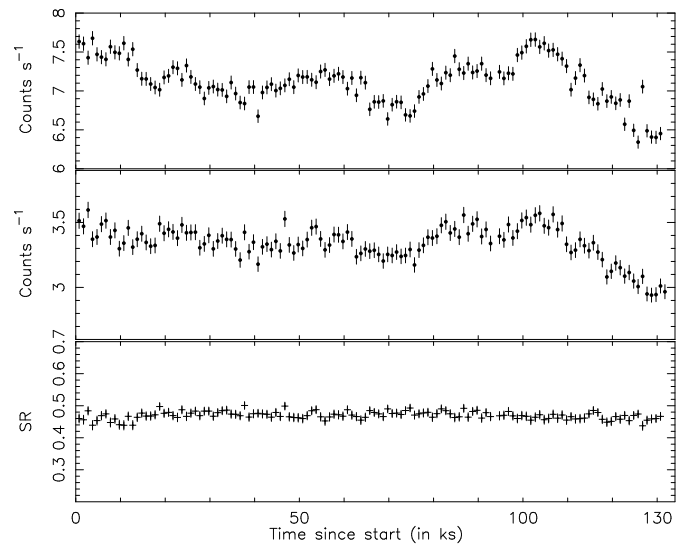}}
\caption{Lightcurves for the pn instrument. Top: the hard band between 2 $-$ 10 keV; middle: the soft band between 0.4 $-$ 1 keV. Bottom: the softness ratio: the 0.4 $-$ 1 keV count rate divided by the 2 $-$ 10 keV count rate. All the data were binned in 1000 s bins.}
\label{fig:hr}
\end{figure}

We constructed the light curve using the pn in the 0.4 $-$ 10 keV band (see Fig.~\ref{fig:pn_light}). Time is measured from the start of the observation, at the 6$^{th}$ of August 2003 at 6$^h$ 13$^m$ 47$^s$ UT. Fig.~\ref{fig:hr} shows the lightcurves in the hard (2 $-$ 10 keV) band, the soft (0.4 $-$ 1 keV) band and the softness ratio determined from these two bands. 
\par
The count rate of the pn varied between 20 counts s$^{-1}$ at the start of the observation and 16 counts s$^{-1}$ at the end of the observation. The most noticeable part of the light curve is the steady decline lasting 25 ks near the end of the observation, after which it levels off for the last 3 ks. Throughout the observation several gradual luminosity changes are observed. 

We identify in Fig.~\ref{fig:pn_light} several parts in the light curve where the total luminosity decreases or increases nearly monotonically. The smallest duration for a 5 \% luminosity variation is $\sim$ 10 ks. The characteristic timescale, $\tau$ = $\mid$d$t$/dln$L$ $\mid$, for these luminosity variations is between 100 and 300 ks. The r.m.s. variance measured as the intrinsic spread for this dataset is 5~\%. From the softness ratio (see Fig.~\ref{fig:hr}) no time lag can be discerned and there is no spectral change with increasing luminosity observed. 

\section{Spectral analysis}
\subsection{Continuum\label{sect:cont}}
\begin{figure}
 \resizebox{\hsize}{!}{\includegraphics[angle=-90]{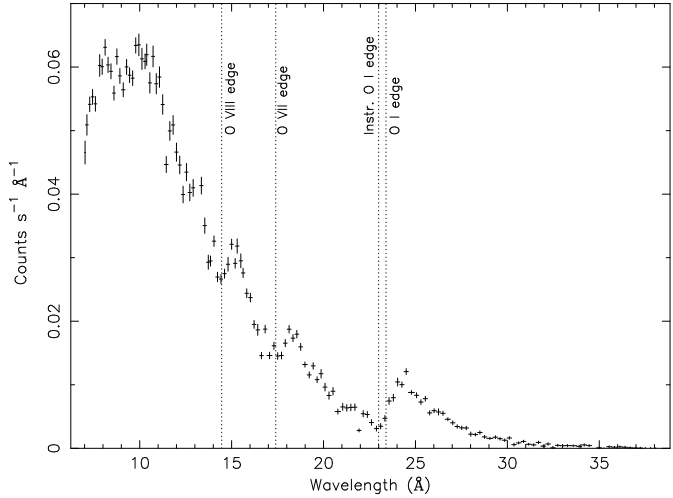}}
\caption{RGS spectrum of IC~4329A with the strongest absorption edges identified. RGS 1 and RGS 2 are averaged for clarity. Note that the edges are convolved with the resolution of the RGS and are heavily blended, thus they are not sharp.}
\label{fig:crgs_tot}
\end{figure}

\begin{figure}
 \resizebox{\hsize}{!}{\includegraphics[angle=-90]{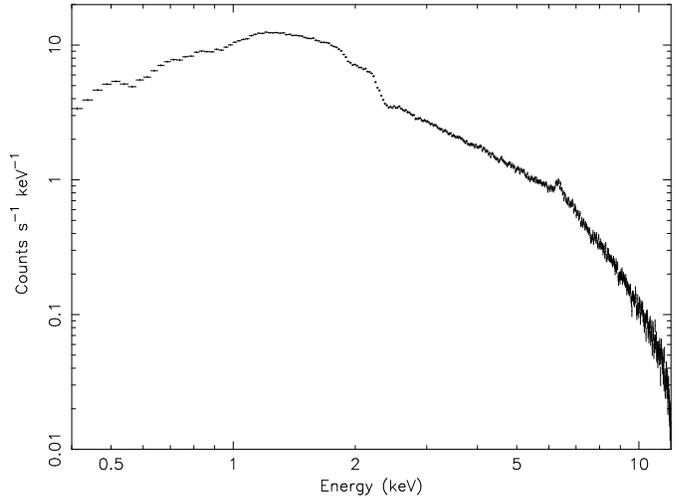}}
\caption{The pn spectrum of IC~4329A. Note the strong absorption at lower energies (see Fig.~\ref{fig:chi_pn} for the residuals to a power-law fit) and the strong Fe K$\alpha$ emission line.}
\label{fig:pn}
\end{figure}

The spectrum is heavily absorbed, with several deep edge-like structures (see Fig.~\ref{fig:crgs_tot} for RGS and \ref{fig:pn} for EPIC-pn). Below energies of about 1.3 keV (i.e. above $\sim$ 10 \AA) we need a complex absorption model (see Sect.~\ref{sect:abs}), a power law continuum and a soft excess which we successfully model with a high temperature modified black body. This is a black body component modified by coherent Compton scattering, for more detail see Kaastra \& Barr (\cite{kaastra89}). Above 1.3 keV the spectrum is well described by a power-law plus two Gaussians for Fe K$\alpha$ and \ion{Fe}{xxvi} Ly$\alpha$ emission line probably blended with the Fe K$\beta$ line (see Sect.~\ref{sect:fek}), corrected for Galactic absorption (see Fig.~\ref{fig:pn} and Fig~\ref{fig:chi_pn}). We modeled the Galactic absorption toward IC~4329A with an absorber in collisional ionization equilibrium, namely the {\it hot} model in SPEX. The {\it hot} model determines the transmission for a certain temperature and total hydrogen column density. This model includes more accurate atomic data for the oxygen edge and includes the neutral oxygen absorption lines. We froze the temperature to 0.5 eV (resulting in an almost neutral gas) and the column density to $N_{\rm H}$ = 4.55 $\times$ 10$^{24}$ m$^{-2}$ (Elvis et al. \cite{elvis89}).  We froze the abundances to solar and the velocity broadening, the Gaussian $\sigma$ to the standard 100 km s$^{-1}$.

In our spectral analysis of the low energy part of the spectrum we focus on the pn and RGS spectra. The pn spectrum is used to determine the continuum and broad emission line parameters, while the RGS spectra are used for fitting the different absorbers in detail. We did not fit the MOS 1 and MOS 2 spectra simultaneously with the pn and RGS spectra as the $\sim$ 10\% systematic uncertainties in the effective areas between the different instruments influence the fit noticeably. The best fit absorption model obtained from the pn and RGS spectra alone matches the MOS spectra rather well, if we allow for different continuum parameters. Table~\ref{tab:tab2} lists the best continuum parameters for the simultaneous pn, RGS 1 and RGS 2 fit. We did not include a continuum reflection component as detected by Gondoin et al. (\cite{gondoin}) from the {\it Beppo}SAX data in our spectral analysis. The XMM-{\it Newton} spectra only extend to about 12 keV, where the continuum reflection component is still negligible (see Fig.~\ref{fig:chi_pn}). 

\begin{table}
\caption{The best fit continuum parameters, for the fit including the absorbers detailed in Sect.~\ref{sect:abs}, fitting the pn and RGS 1 and RGS 2 simultaneously.}
\label{tab:tab2}
\begin{center}
\begin{tabular}{|l|l|l|} \hline
pl    & norm.$^{a}$& 2.90 $\times$ 10$^{52}$ ph s$^{-1}$ keV$^{-1}$ \\
      & lum.$^{a,b}$ & 1.17 $\times$ 10$^{37}$ W \\
      & $\Gamma$$^{a}$ & 1.709  \\
mbb   & norm.$^{a}$& 7.2 $\times$ 10$^{31}$ m$^{1/2}$ \\
      & $kT$$^{a}$  & 473 eV \\\hline
\end{tabular} \\
$^{a}$ The systematic errors dominate over the statistical errors, which are on the order of 1\%. \\
$^{b}$ in the 2 $-$ 10 keV band, calculated from the measured normalization, photon index. 
\end{center}
\end{table}

\subsection{Fe K$\alpha$ and \ion{Fe}{xxvi} Ly$\alpha$ line\label{sect:fek}}
\begin{figure}
 \resizebox{\hsize}{!}{\includegraphics[angle=-90]{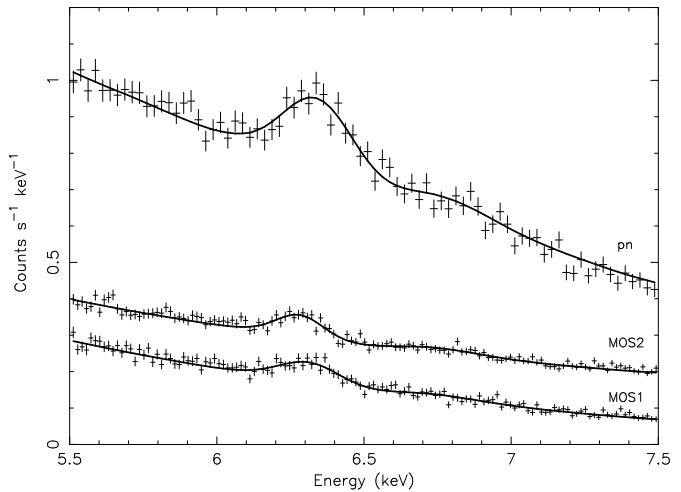}}
\caption{Detail of the pn, MOS 1 and MOS 2 spectra showing the Fe K$\alpha$ and \ion{Fe}{xxvi} Ly$\alpha$ emission lines. The model through the data is given in Table~\ref{tab:tab3}. The MOS 2 data were shifted by adding 0.006 counts s$^{-1}$ for clarity.}
\label{fig:pnfe}
\end{figure}

\begin{figure}
 \resizebox{\hsize}{!}{\includegraphics[angle=-90]{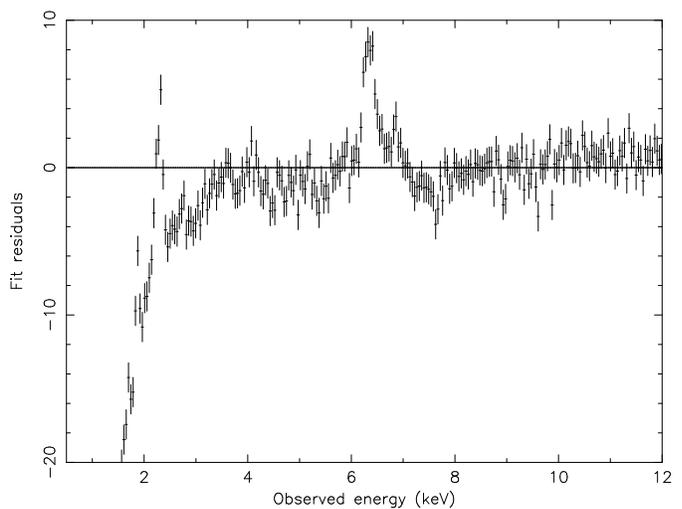}}
\caption{A power-law plus Galactic absorption fit to the pn spectrum, rebinned by a factor of 9. Note the Fe K$\alpha$ and \ion{Fe}{xxvi} Ly$\alpha$ emission lines, as well as the absorption at lower energies. The feature at 2 keV is an artifact of the absorption spectrum and the soft excess. Note that the reflection component has at most a 1 $\sigma$ significance.}
\label{fig:chi_pn}
\end{figure}

A narrow Fe K$\alpha$ line is clearly detected in all three EPIC spectra. There is a second, weaker line detected at about 6.9 keV, which we identify as a blend of the 6.95 keV and 6.97 keV \ion{Fe}{xxvi} Ly$\alpha$ lines and probably blended with the Fe K$\beta$ line at 7.06 keV (Palmeri et al. \cite{palmeri}). In Fig.~\ref{fig:pnfe} the pn, MOS 1 and MOS 2 spectra around the Fe K$\alpha$ line are plotted, in Fig.~\ref{fig:chi_pn} the residuals to a power-law fit of the pn spectrum is shown. 
\par
We tested the accuracy of the energy scale for all instruments by looking at instrumental background lines, to avoid any gain problem. For the MOS cameras the outer CCDs are exposed, and we used the whole exposed area to find the instrumental background lines. For MOS~1 the strongest instrumental line, Al at 1.4866 keV is detected at the expected energy. For MOS~2 all the background lines, with the exception of the weak Ni line at 7.4724 keV are detected at the expected energy. Therefore the energy scale of the MOS spectra should be correct. For pn we are limited to CCD 4; however, we do not detect any of the the instrumental background lines with sufficient significance to be able to verify the energy scale. There can thus still be a small gain problem for the pn. 
\par
Table~\ref{tab:tab3} lists the best fit values fitting Gaussian profiles to both iron emission lines in the pn and the MOS spectra. The full width half maximum (FWHM) and energy of the Fe K$\alpha$ line are consistent within the instrumental uncertainties. For the line at 6.9 keV data for the pn and the MOS instruments are consistent. Interestingly, we do not detect the \ion{Fe}{xxv} triplet around 6.70 keV. 

\begin{table}
\caption{The best fit parameters for the Fe K$\alpha$ and \ion{Fe}{xxvi} Ly$\alpha$ emission lines for the pn and MOS 1 and MOS 2 spectra, using Gaussian profiles. As the line at $\sim$6.9 keV is rather weak, we fitted MOS 1 and MOS 2 simultaneously. The energies quoted are for the restframe of IC~4329A. The errors on the line energies are the statistical uncertainties since these are larger than the systematic uncertainties in the energy scale. All quoted results have been corrected for the redshift of IC~4329A.}
\label{tab:tab3}
\begin{center}
\begin{tabular}{|l|c|c|c|} \hline
                  & pn              & MOS1           & MOS2  \\\hline
 norm$^{a}$ & 9.5 $\pm$ 0.9   & 7.6 $\pm$ 0.9  & 6.4 $\pm$ 0.8 \\
 flux$^{b}$ & 0.85 $\pm$ 0.08 & 0.66 $\pm$ 0.08& 0.58 $\pm$ 0.07\\
 $E$ (keV)  & 6.45 $\pm$ 0.01 & 6.42 $\pm$ 0.01& 6.40 $\pm$ 0.01\\
 FWHM (eV)  & 240 $\pm$ 50    & 110 $\pm$ 50   & $<$ 100  \\
 EW (eV)    & 87 $\pm$ 8      & 68 $\pm$ 8     & 64 $\pm$ 8 \\\hline
 norm$^{a}$ & 3.0 $\pm$ 0.8   & \multicolumn{2}{c|}{3.2 $\pm$ 0.9} \\
 flux$^{b}$ & 0.27 $\pm$ 0.07 & \multicolumn{2}{c|}{0.28 $\pm$ 0.08}\\
 $E$ (keV)  & 6.92 $\pm$ 0.05 & \multicolumn{2}{c|}{6.89 $\pm$ 0.05}\\
 FWHM (eV)  & 300 $\pm$ 100   & \multicolumn{2}{c|}{300 $\pm$ 100}\\
 EW (eV)    & 30 $\pm$ 8      & \multicolumn{2}{c|}{36 $\pm$ 10}  \\\hline
\end{tabular} \\
$^{a}$ in $\times$ 10$^{49}$ ph s$^{-1}$. \\
$^{b}$ in ph m$^{-2}$ s$^{-1}$ 
\end{center}
\end{table}

\begin{figure}
 \resizebox{\hsize}{!}{\includegraphics[angle=-90]{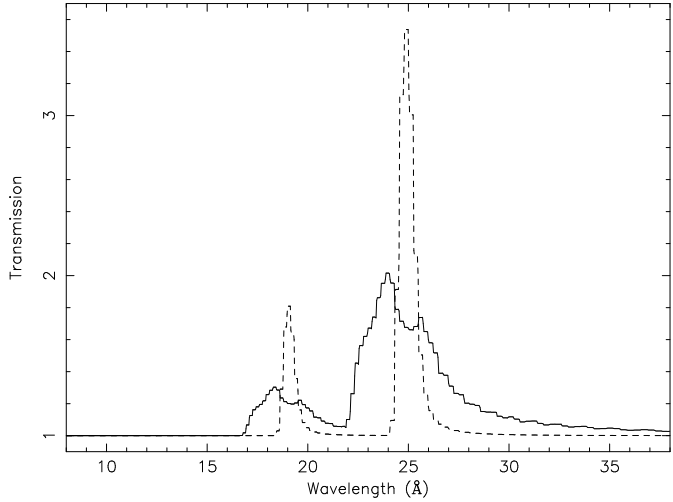}}
\caption{Models for the two broadened emission features detected in the RGS spectrum (see Sect~\ref{sect:bemission}) and the two emission features detected in the pn spectrum. The thin solid line represents the best fit for the relativistically broadened lines detected in the RGS spectra, with $q$ = 1.9 and $i$ = 80.9 (see Table~\ref{tab:tab5}). The dashed line represents the model for the emission features adopting the parameters found for the relativistically broadened Fe K$\alpha$ and \ion{Fe}{xxvi} emission lines: emissivity slope $q$ = 1 and $i$ = 12$^{\circ}$, consistent with the values obtained by McKernan \& Yaqoob (\cite{mckernan}). Note that this profile is very similar to a Gaussian profile.}
\label{fig:laor}
\end{figure}

Gondoin et al. (\cite{gondoin}) find an absorption edge at 7.1 keV with an depth of 0.03 $\pm$ 0.02. There is a 1 $\sigma$ jump in our data at 7.1 keV, corresponding to an optical depth of $<$ 0.04. However, we do not detect sufficient iron L line absorption between 15.5 \AA~and 16.5 \AA~and edge absorption between 8 \AA~and 10 \AA~in the RGS spectra to explain such an edge. As a result we decided against the interpretation that this feature is an edge. McKernan \& Yaqoob (\cite{mckernan}) note two emission features in the {\it Chandra} HEG spectrum, which they model with either two Gaussians or two disklines. In the model with two Gaussians they obtain for the Fe K$\alpha$ line an energy of 6.30 $\pm$ 0.08 keV, for the \ion{Fe}{xxvi} Ly$\alpha$ line an energy of 6.91 $\pm$ 0.04 keV. Both are consistent with our results, as are the measured EWs and intensities. We also tested a double diskline model (Laor \cite{laor}), which gives a similarly good fit, for two more free parameters as compared to the fit with two Gaussians. For pn we find an inclination angle $i$ $<$ 17$^{\circ}$ and an emissivity slope $q$ $<$ 1.4, where for large radii $I(r)$ $\sim$ $r^{-q}$. This is consistent with the parameters determined by McKernan \& Yaqoob (\cite{mckernan}) who find an inclination angle of 12$^{\circ}$ $\pm$ 1$^{\circ}$ and a flat emissivity slope. For the MOS spectra we find similar, but less constrained upper limits. The diskline profile for $i$ = 12$^{\circ}$ and $q$ = 1 is shown in Fig.~\ref{fig:laor} (dashed line). Note that this profile is very similar to a Gaussian line profile. As a result, the pn and MOS spectra do not allow us to distinguish between a model with two disklines or two Gaussian emission lines. As we cannot constrain the diskline parameters, we use the model with two Gaussians in the following data analysis.
\par
We re-analyzed the pn spectrum for the 2001 XMM-{\it Newton} observation of IC~4329A. We used SAS version 6.0 and the PPS data files for the data reduction. The data are noisy (total exposure time of 9.7 ks), leading to large error bars or upper limits for the Fe K$\alpha$ parameters, even if we fit the emission line with a Gaussian. In Table~\ref{tab:tabx} we compare the different literature values for the Fe K$\alpha$ emission line. Note that the flux of the Fe K$\alpha$ line is consistent with a constant value since August 1997, although there is a rather large spread in FWHM. However, the error bars on the flux values are large, and thus we can not exclude variability. The large spread in FWHM is partially a result from differing spectral resolution of the different instruments. An indication of the uncertainty in absolute calibration between the different instruments are the differing parameters for the simultaneous ASCA and RXTE observations (Done et al. \cite{done}).

\begin{table*}
\caption{Comparison of the Fe K$\alpha$ emission line as observed in earlier observations.}
\label{tab:tabx}
\begin{center}
\begin{tabular}{|l|lllll|} \hline
             & ASCA$^{a}$ & RXTE$^{a}$& BeppoSAX$^{b}$ & XMM-Newton$^{c}$ & Chandra$^{d}$\\
             & 1997 Aug. 7-16 & 1997 Aug. 7-16 & 1998 Jan. 02 & 2001 Jan. 31 & 2001 Aug. 26 \\\hline
Energy (keV) & 6.37$\pm$0.06 & 6.3$\pm$0.1 & 6.5$\pm$0.2    &  6.5$\pm$0.2 & 6.30$\pm$0.08\\
FWHM (eV)    & 920$\pm$240   & 1150$\pm$450 & 850$\pm$590    &  410$\pm$500    & 440$\pm$190 \\
flux$^{e}$   & 0.7$\pm$0.2$^{f}$ & 1.2$\pm$0.2$^{f}$ & 1.6$\pm$0.7 & $<$0.9$^{g}$ & 1.9$\pm$0.8 \\
EW (eV)      & 180$\pm$50    & 240$\pm$45  & 109$\pm$44     &  $<$72$^{g}$     & 110$\pm$43\\\hline
\end{tabular} \\
$^{a}$ Done et al. (\cite{done}), the ASCA and RXTE data were taken simultaneous.\\
$^{b}$ Perola et al. (\cite{perola}).\\
$^{c}$ based on our own analysis of the pn data. \\
$^{d}$ McKernan \& Yaqoob (\cite{mckernan}) for the fit with two Gaussian lines.\\
$^{e}$ in ph m$^{-2}$ s$^{-1}$. \\
$^{f}$ Calculated from the given EW and continuum level. \\
$^{g}$ The 3 $\sigma$ upper limit.
\end{center}
\end{table*}

\section{Absorbers\label{sect:abs}}

\begin{figure}
 \resizebox{\hsize}{!}{\includegraphics[angle=-90]{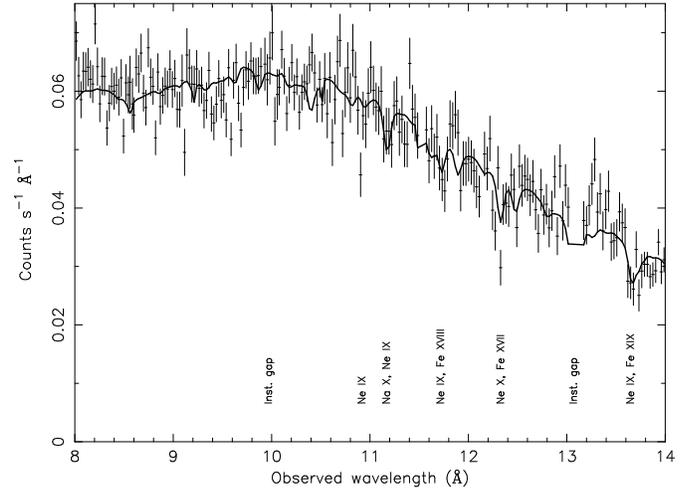}}
\caption{Detail of the RGS spectrum between 8 and 14 \AA. The solid line through the data points is our best fit model as described in the text. RGS 1 and 2 are added for clarity in this and the next two spectra.}
\label{fig:rgs_0814}
\end{figure}

\begin{figure}
 \resizebox{\hsize}{!}{\includegraphics[angle=-90]{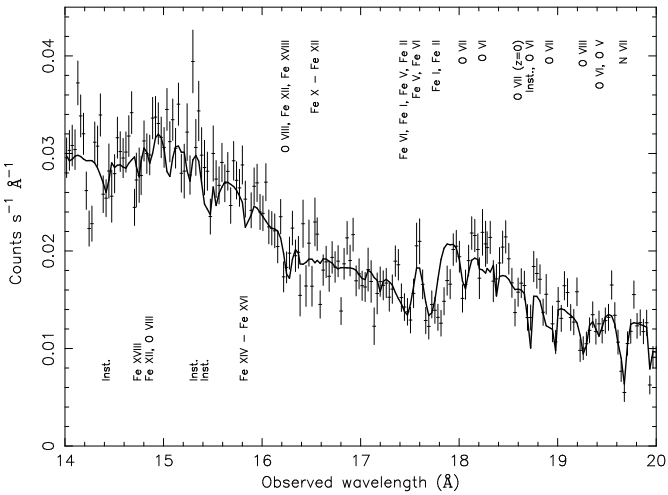}}
\caption{Detail of the RGS spectrum between 14 and 20 \AA. The solid line through the data points is our best fit model as described in the text. The \ion{O}{vii} absorption at z = 0 is indicated, but not fit in the plot.}
\label{fig:rgs_1420}
\end{figure}

\begin{figure}
 \resizebox{\hsize}{!}{\includegraphics[angle=-90]{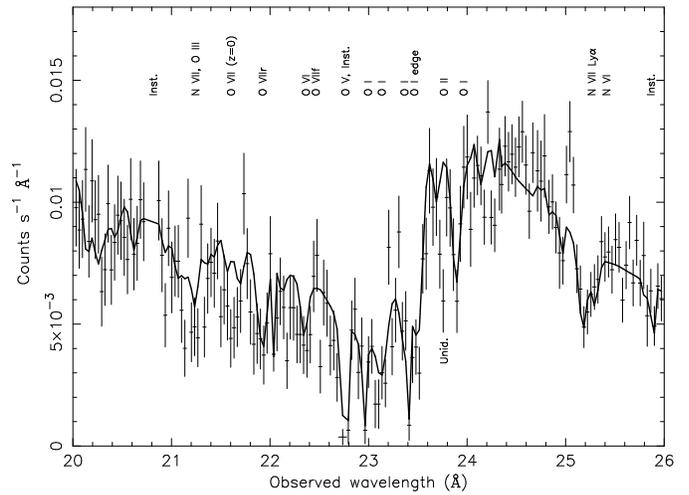}}
\caption{Detail of the RGS spectrum between 20 and 26 \AA. The solid line through the data points is our best fit model as described in the text. The \ion{O}{vii} absorption line at z = 0 and the \ion{O}{vii} forbidden line are indicated, but not fit in the plot. The feature labeled as Unid. is further discussed in the text.}
\label{fig:rgs_2026}
\end{figure}

The RGS spectra are dominated by absorption edges (see Fig.~\ref{fig:crgs_tot}), as was also concluded by Gondoin et al. (\cite{gondoin}). The earlier 12~ks RGS observation is too short to include useful high resolution spectral information. As the warm absorber could be variable on timescales much shorter than the time difference between both observations, we did not use the earlier RGS data in our analysis. There are only a few absorption lines that can be easily identified. We model the absorber using a combination of {\it hot} and {\it xabs} components. The {\it hot} model calculates the transmission (both lines and continuum) of a thin slab in collisional ionization equilibrium. The {\it xabs} model does the same for a photoionized model. In both models the total column density, outflow velocity $v$ and velocity dispersion (Gaussian $\sigma$) $\sigma_v$ are free parameters, as well as the temperature $T$ for the {\it hot} model and the ionization parameter $\xi$ = $L$/$nr^2$ for the {\it xabs} model. For more details see Kaastra et al. (\cite{kaastra02}; \cite{kaastrasp}). To fit the spectrum self-consistently we need at least seven absorbers (see Table~\ref{tab:tab4}). The best fit spectrum including these absorption components is shown in Fig~\ref{fig:rgs_0814} $-$~\ref{fig:rgs_2026}. Component 1 is the absorption due to our own Galaxy discussed in Sect.~\ref{sect:cont}. The absorber with the largest opacity is neutral and due to the host galaxy of IC~4329A, component 2, and is easily identified from its redshifted \ion{O}{i} absorption line (see Fig.~\ref{fig:rgs_2026}). Fitting this absorber improves the $\chi^2$ from 100398 to 16872 for 1603 degrees of freedom (dof) for the continuum parameters quoted in Table~\ref{tab:tab2}. This absorber is discussed in Sect.~\ref{sect:host}. 

The absorber with the highest column density, component 5, is highly ionized, and including this absorber further reduces $\chi^2$ to 13559 for 1607 dof. This component is responsible for the \ion{O}{vii} and \ion{Ne}{ix} resonance lines, the \ion{N}{vii} Ly$\alpha$ and \ion{O}{viii} Ly$\alpha$ lines; as well as absorption edges due to these ions. Further, a less ionized absorber, component 4 with also a lower column density is necessary to fit the \ion{O}{v} $-$ \ion{O}{vii} lines and some iron M shell Unresolved Transition Array (UTA) (Behar, Sako \& Kahn \cite{behar}). Adding component 4 reduces the $\chi^2$ to 11947 for 1611 dof. A very highly ionized absorber could also be present (component 6), but is only weakly detected, and only improves the $\chi^2$ by 507 for 4 extra degrees of freedom. This component produces (a part of) \ion{O}{viii} and \ion{Ne}{x} Ly$\alpha$ lines. The fit improves significantly, reducing $\chi^2$ to 2428 for 1615 dof, by adding a near neutral absorber (component 3). This fits a part of the UTA as well as absorption from \ion{O}{ii}. However, this component is not significant if we assume that the gas in the host galaxy is lowly ionized. To distinguish between a neutral or a cold absorber the detection of \ion{O}{ii} is crucial.  All these absorbers are further discussed in Sect.~\ref{sect:warm}. 
\par
Finally, we detect a weak \ion{O}{vii} absorption component (7) from $z$ = 0 plasma, which is discussed in Sect.~\ref{sect:emission}.

\subsection{Absorption from the host galaxy: component 2\label{sect:host}}
This is the absorption component also detected in the optical and is responsible for the reddening observed in the optical and the UV. The X-ray \ion{O}{i} absorption lines are well explained by a neutral absorber. This absorption system is the dominant component (2), in the sense of the opacity in the X-ray spectrum. 
\par
\begin{figure}
 \resizebox{\hsize}{!}{\includegraphics[angle=-90]{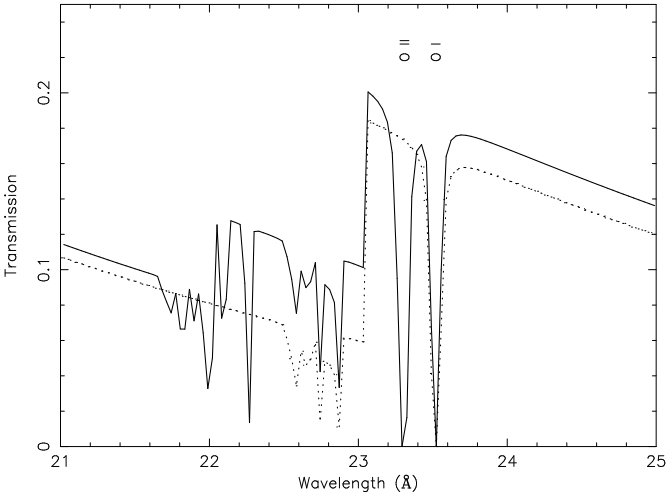}}
\caption{The difference in transmission for the near neutral absorber ($T$ = 1.2 eV, thin line) and a neutral absorber ($T$ = 0.5 eV, dotted line) for the wavelength range between 21 and 25 \AA. The absorption lines in both models are due to \ion{O}{i}. The absorption lines only detected for the near neutral absorber are due to \ion{O}{ii}; note the deep \ion{O}{ii} line at 23.3 \AA.}
\label{fig:hotdetail}
\end{figure}

To ascertain that this absorber is neutral and not lowly ionized, we plot the transmission for a neutral and for a lowly ionized absorber in Fig.~\ref{fig:hotdetail}. The absorption lines from \ion{O}{ii} around 22~\AA~and the deep \ion{O}{ii} absorption line at about 23.3 \AA~are the only detectable differences between these transmission models. Juett, Schulz \& Chakrabarty (\cite{juett}) determined from a study of the interstellar medium with the {\it Chandra} HETG that the \ion{O}{ii} line wavelength is 23.33~\AA. In the RGS spectrum of IC~4329A there is a deep line at 23.37 \AA~$\pm$ 0.01~\AA~rest wavelength (23.74~\AA~observed wavelength, identified as Unid. in Fig.~\ref{fig:rgs_2026}) for which there is no straightforward identification. The 0.04 \AA~difference is larger than the 8 m\AA~uncertainty in the absolute calibration of the RGS, and is inconsistent with the fact that we detect \ion{O}{i} at the correct wavelength. 
\par
If this line at 23.37 \AA~is identified as an \ion{O}{ii} line, then the absorber of the host galaxy should be redshifted by 500 km s$^{-1}$ and not neutral but very lowly ionized. Inflow is not expected if the gas is part of the host galaxy. Therefore we conclude that component 2 does not contain a significant amount of \ion{O}{ii} and in the further analysis we will assume that this absorber is neutral.

\begin{table*}
\caption{Absorption components in the IC~4329A spectrum as determined from the RGS spectra. The second order was also included in the fit, so as to have a more accurate measurement of the velocity width as well as the outflow velocity. The outflow velocity was frozen to 0 km s$^{-1}$ for the component 2 and 3 as some of the stronger lines in the spectrum have uncertain wavelength, or are detected only in the form of blends.}
\label{tab:tab4}
\begin{center}
\begin{tabular}{|l|ccccccc|} \hline
comp.            & 1         & 2               & 3                  & 4              & 5              & 6           & 7  \\
origin           & Gal.      & IC~4329A        & IC~4329A           & IC~4329A       & IC~4329A       & IC~4329A    & Gal/ISM\\
model            & {\it hot} & {\it hot}       & {\it xabs}         & {\it xabs}     & {\it xabs}     & {\it xabs}  & see text \\\hline
$N_{\rm H}^{a}$  & 0.455$^{e}$& 1.732 $\pm$ 0.009& 1.32 $\pm$ 0.03  & 0.32 $\pm$ 0.03 & 6.6 $\pm$ 0.4  & 2.0 $\pm$ 0.5 & $-$\\
log $\xi^{b}$    & $-$       & $-$             & $-$1.37 $\pm$ 0.06 & 0.56 $\pm$ 0.10& 1.92 $\pm$ 0.03& 2.70 $\pm$ 0.06 & $-$\\
$T^{c}$          & 0.5$^{e}$ & 0.5$^{e}$       &  $-$               & $-$            & $-$            & $-$             & $-$\\
$v^{d}$         & 0$^{e}$    & 0$^{e}$         & 0$^{e}$            & $-$200 $\pm$ 100& $-$100 $\pm$ 100& 20 $\pm$ 160  & $-$ \\
$\sigma_v^{d}$  & 50$^{e}$   & $<$ 50          & 45 $\pm$ 20        & 90 $\pm$ 60    & 70 $\pm$ 10    & 140 $\pm$ 75    & $-$\\\hline
\end{tabular} \\
$^{a}$ in 10$^{25}$ m$^{-2}$. \\
$^{b}$ in 10$^{-9}$ W m. \\
$^{c}$ in eV. \\
$^{d}$ in km s$^{-1}$. \\
$^{e}$ fixed parameter. \\
\end{center}
\end{table*}

\subsection{Warm absorbers\label{sect:warm}}

To explain most of the strongest lines we need an ionized absorber. The dominant component 5 is modeled by a {\it xabs} component and has an ionization parameter log $\xi$ = 1.92, see Table~\ref{tab:tab4}. $\xi$ is measured in units of 10$^{-9}$ W m. Due to the low signal-to-noise ratio we froze the elemental abundances to solar while fitting the spectrum. This absorber fits the \ion{O}{vii} and \ion{Ne}{ix} resonance lines as well as the \ion{N}{vii} Ly$\alpha$ and \ion{O}{viii} Ly$\alpha$ and (a part of) some weaker absorption lines such as \ion{Ne}{x} and highly ionized iron. There is thus a warm absorber in IC~4329A similar to those detected in other narrow-line Seyfert 1 galaxies. The total hydrogen column density detected for this absorber ranges from 1.32 $\times$ 10$^{25}$ m$^{-2}$ for the lowest ionized absorber component 3 to 6.6 $\times$ 10$^{25}$ m$^{-2}$ for component 5. These total hydrogen column densities are similar to the hydrogen column density detected in other Seyfert 1 galaxies such as NGC~5548 (Kaastra et al. \cite{kaastra02}; Steenbrugge et al. \cite{steen}).
\par
\begin{figure}
 \resizebox{\hsize}{!}{\includegraphics[angle=-90]{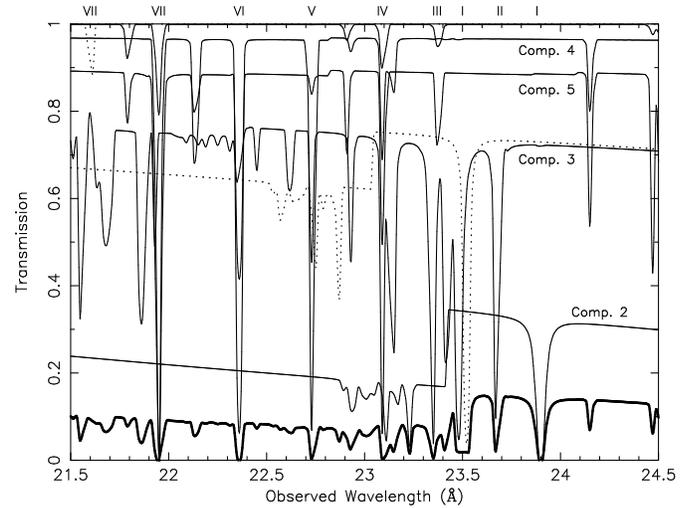}}
\caption{Part of the spectrum detailing the transmission for the seven different absorption components. The thick line indicates the best fit spectral model, i. e. with all seven absorption models applied. The dotted line at the top with only an \ion{O}{vii} absorption line is component 7, the other dotted line is component 1, or Galactic absorption. The solid line on the top represent component 6, the other components are labeled in the plot. In roman numerals the ionization state of oxygen is indicated, from neutral oxygen (I) to \ion{O}{vii} (VII).}
\label{fig:abs4329a}
\end{figure}

There are some remaining absorption features in the spectrum that can be fit by a relatively cool absorber (component 3, log $\xi$ = $-$1.37), which in particular produces lines and an edge of \ion{O}{iii} as well as some continuum depression due to the UTA of iron. We modeled this absorber with a {\it xabs} model. To test that this lowly ionized absorber is not part of the host galaxy we tried fitting this absorber with a {\it hot} component. This worsens $\chi^2$ by 21 for 2377 dof, letting the other absorbers free in the fit. However, as the difference in the model is mainly the continuum depression due to the UTA, no strong conclusions can be drawn as to whether this absorber is photoionized and is part of the warm absorber or is collisionally ionized and belongs to the host galaxy. Fig.~\ref{fig:abs4329a} details the models of these seven absorption components and the net absorption model.
\par

\begin{table}
\caption{The model column densities of the most prominent ions in the spectrum of IC~4329A. For each ion we indicated the absorption component that is dominant in producing absorption from this ion.}
\label{tab:taby}
\begin{center}
\begin{tabular}{|lcr|lcr|} \hline
Ion           &  log N$_{\rm H}$ & comp  &Ion           &  log N$_{\rm H}$ & comp  \\
              &  (m$^2$)   &       &              &  (m$^2$)   & \\\hline
\ion{N}{vii}  &  22.1      & 5     & \ion{Ne}{x}  &  21.6  & 5 \\
\ion{O}{i}    &  22.2      & 2     & \ion{Fe}{iv} &  20.0  & 3\\
\ion{O}{ii}   &  20.6      & 3     & \ion{Fe}{vi} &  20.1  & 3\\
\ion{O}{iii}  &  21.8      & 3     & \ion{Fe}{xviii}& 20.0 & 5\\
\ion{O}{iv}   &  21.6      & 3     & \ion{Fe}{xix}& 20.0   & 6 \\
\ion{O}{v}    &  20.9      & 3,4   & \ion{Fe}{xx} & 20.2   & 6\\
\ion{O}{vi}   &  21.0      & 4     & \ion{Fe}{xxi}& 20.2   & 6\\
\ion{O}{vii}  &  21.7      & 5     & \ion{Fe}{xxii}& 20.4  & 6\\
\ion{O}{viii} &  22.4      & 5     & \ion{Fe}{xxiii}& 20.3 & 6\\
\ion{Ne}{ix}  &  21.4      & 5     &                &      &  \\\hline
\end{tabular}
\end{center}
\end{table}

In most warm absorbers known up to now, a range of ionization parameters is detected. Adding component 4 (log $\xi$ = 0.56) improves $\chi^2$ by 52 for 2373 dof. Finally, we detect a highly ionized absorber with log $\xi$ = 2.70, but with a small total hydrogen column density. 
\par
No strong narrow emission lines are detected in the soft part of the spectrum. We found an  upper limit of 1 $\times$ 10$^{50}$ photons s$^{-1}$  or an EW $<$ 0.6 \AA~for the \ion{O}{vii} forbidden line; and 1.4 $\times$ 10$^{49}$ photons s$^{-1}$ or an EW $<$ 0.03 \AA~for the \ion{Ne}{ix} forbidden line. There are 1 $\sigma$ significant emission features at 8.54~\AA~and 11.65~\AA, however, as there is no straightforward identification for these features these are probably due to hot pixels or noise.

\begin{figure}
 \resizebox{\hsize}{!}{\includegraphics[angle=-90]{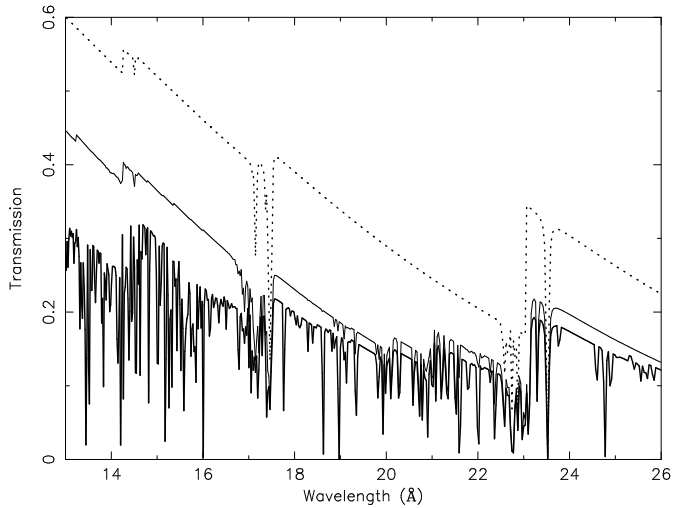}}
\caption{The transmission of the neutral absorber (dotted line), the neutral and lowly ionized absorber (thin line) and the transmission also including the medium and highly ionized absorbers (thick line).}
\label{fig:hotxabs12}
\end{figure}

\subsection{Absorption at z = 0: component 7\label{sect:emission}}
The last absorbing system detected is due to hot gas at $z$ = 0, previously detected in the spectra of other AGN (Rasmussen et al. \cite{rasmussen}; Nicastro et al. \cite{nicastro}; Steenbrugge et al. \cite{steen}; Behar et al. \cite{behar1}). In the IC~4329A spectrum there are two absorption lines due to this local gas, i.e. with zero redshift. We identified both as due to \ion{O}{vii}, the strongest line being the resonance line at 21.6 \AA~(see Fig.~\ref{fig:rgs_2026}). For this line we measure an equivalent width of 0.03 $\pm$ 0.02 \AA. The absorption line at 18.62 \AA~(see Fig.~\ref{fig:rgs_1420}) is expected to be weaker and has an EW of 0.01 $\pm$ 0.01 \AA. Interestingly, no \ion{O}{viii}~Ly$\alpha$ line is detected. The \ion{O}{vi} line is too blended with a redshifted \ion{O}{i} absorption line to be detectable. The \ion{O}{v} line at 22.374 \AA~is also not detected at $z$ = 0. Assuming that the absorbing gas is in collisional equilibrium, we determine provisionally a temperature range between 0.03 and 0.2 keV. However, from detecting only two \ion{O}{vii} lines we cannot conclude that the gas is in collisional equilibrium. Assuming that the gas is photoionized, which is less likely as the source of radiation is not known, we find an ionization parameter of log $\xi$ $\sim$ 0.8. The presence of these absorption lines in the spectrum indicates that using AGN spectra it is possible to obtain a measure for the amount of highly ionized local gas, as well as an indication of its temperature or ionization balance. 

\subsection{Broad emission lines\label{sect:bemission}}
\begin{figure}
 \resizebox{\hsize}{!}{\includegraphics[angle=-90]{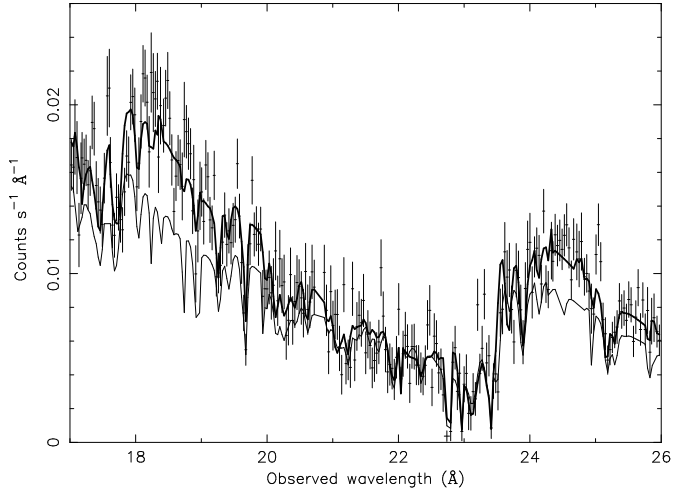}}
\caption{The best fit model with the disklines fit (thick line) versus the model without these excesses fit (thin line).}
\label{fig:brcomp}
\end{figure}

Compared to the best fit with our absorption model, there appears to be an excess at the wavelengths around \ion{N}{vii} Ly$\alpha$ and \ion{O}{viii}~Ly$\alpha$ in both the RGS and pn spectra. These excesses can be fit by either broad Gaussians or by diskline profiles. For the diskline profile we used one set of parameters for the inner and outer edge of the accretion disk, the emissivity slope $q$ (as defined in Sect.~\ref{sect:fek}) and the inclination angle $i$. The result of the model with and without these diskline profiles can be seen in Fig.~\ref{fig:brcomp}. Leaving the wavelength of the lines frozen to the rest wavelength in IC4329A, the diskline model gives a substantial better fit, lowering $\chi^2$ by 368 for 2367 dof. However even in the fit with the disklines, there is still an excess, most pronounced at shorter wavelengths. For the \ion{N}{vii} Ly$\alpha$ emission line this excess produces a narrow emission line like feature at 25.05 \AA.  

Table~\ref{tab:tab5} compares the model with Gaussian and disklines profiles.  Adding the disklines reduces the modified black body temperature, as the excess is partly fitted by this continuum component. However, the newly derived temperature for the modified black body is still quite high, compared to other Seyfert 1 galaxies. We did not include a \ion{C}{vi} Ly$\alpha$ emission line as the absorption at the longer wavelength is too strong. Note that the inclination determined for the disklines is rather high, similar to the inclination of the host galaxy. This is in contrast to the inclination determined from the hard X-ray Fe K$\alpha$ and \ion{Fe}{xxvi} Ly$\alpha$ emission lines, which however are also very well fit with narrow Gaussians.

\begin{table}
\caption{Best fit results for the emission lines observed in the soft X-ray part of the spectrum. The lines were fitted using Gaussians (columns 2 and 3) and disklines (column 4 and 5), in both cases the energy of the line was frozen to its rest-energy at the redshift of IC~4329A. The disklines parameters were fit for both disklines simultaneously.}
\label{tab:tab5}
\begin{center}
\begin{tabular}{|l|ll|ll|} \hline
           & Gaussian    & Gaussian      & diskline       & diskline \\\hline
norm$^{a}$ & 25 $\pm$ 1  & 1.4 $\pm$ 0.2 & 20.4 $\pm$ 0.5 & 4.2 $\pm$ 0.3 \\
$\lambda^{b}$& 24.78   & 18.97         & 24.78          & 18.97   \\
FWHM$^{c}$ & 4.5 $\pm$ 0.1 & 4.1 $\pm$ 0.3 &             &         \\
EW (\AA)   & 5.8 $\pm$ 0.2 & 0.57 $\pm$ 0.08 & 3.8 $\pm$ 0.1 & 0.26 $\pm$ 0.02 \\\hline
$R_{\rm in}^{c}$ &       &               & \multicolumn{2}{c|}{1.3} \\
$R_{\rm out}^{c}$ &      &               & \multicolumn{2}{c|}{400}   \\
$q$        &             &               & \multicolumn{2}{c|}{1.9 $\pm$ 0.4 } \\
$i$ (deg)  &             &               & \multicolumn{2}{c|}{80.9 $\pm$ 0.4 } \\\hline 
\end{tabular} \\
$^{a}$ in 10$^{51}$ photons s$^{-1}$. \\
$^{b}$ in \AA~and in the restframe. \\
$^{c}$ in \AA. \\
$^{d}$ in $GM/c^2$. \\
\end{center}
\end{table}

\section{Discussion\label{sect:disc}}
We detect no variability in the softness ratio, although the luminosity varied by about 17~\%. Perola et al. (\cite{perola}) observed in their broad band BeppoSAX spectrum significant variations in the lightcurve between 0.1 $-$ 100 keV. However, only marginal evidence for variations in the  hardness ratios were detected. Singh, Rao \& Vahia (\cite{singh}) found from EXOSAT data that IC~4329A was continuously variable, with one 12\% change in luminosity over a 20 ks period, corresponding to a characteristic timescale $\tau$ = 170 ks. This characteristic timescale is very similar to those measured from our lightcurve. They did not record variability on timescales shorter than $\sim$ 10 ks. 
\par
In the IC~4329A spectrum there are seven different absorbing systems, and disentangling them is rather complicated. First there is absorption due to cold gas in our own Galaxy (component 1) and warm gas in its surrounding (component 7). On the other hand there is strong absorption due to neutral gas in the host galaxy, as determined from the reddening measured in the optical, UV bands and our X-ray spectrum (component 2). This is consistent with the dusty luke-warm absorbers measured in NGC~3227 and Ark~564 (Crenshaw \& Kraemer \cite{crenshaw}). Further, there is the warm absorber, probably originating close to the central source. This absorber is photoionized and represented by components 3 $-$ 6. 
\par
This warm absorber has a range in ionization parameters and total hydrogen column density similar to NGC~5548 (Kaastra et al. \cite{kaastra02}; Steenbrugge et al. \cite{steen}), NGC~3783 (Blustin et al. \cite{blustin}; Kaspi et al. \cite{kaspi}), and NGC~7469 (Blustin et al. \cite{blustin1}). As such, the warm absorber seems very similar to other sources observed, with an increase in column density with ionization parameter. The outflow velocity measured is similar to the lowest outflow velocity (component 5) of NGC~5548 (Crenshaw et al. \cite{crenshaw03}), and is smaller than the outflow velocity observed in NGC~3783 (Blustin et al. \cite{blustin}). From Table~\ref{tab:tab4} there is a slight trend for increasing velocity dispersion with increasing ionization parameter from $<$ 50 km s$^{-1}$ for lowly ionized component 3 to 140 $\pm$ 75 km s$^{-1}$ for the highest ionized component 6. No trend of velocity dispersion versus ionization parameter was observed for the high signal to noise ratio LETGS observations of NGC~5548 (Steenbrugge et al. \cite{steen}) or other observed Seyfert 1 galaxies. The opposite trend was observed in the Seyfert 2 galaxy NGC~1068 by Brinkman et al. (\cite{brinkman}) where the lowest ionization parameters had the largest velocity dispersion.
\par
There is a broad excess measured at some wavelength ranges, most notably near the \ion{O}{viii}~Ly$\alpha$ (18.969 \AA) and \ion{N}{vii} Ly$\alpha$ (24.781 \AA) emission lines. These are best fit with disklines (solid line in Fig.~\ref{fig:laor}, and the thick line in Fig.~\ref{fig:brcomp}), although even then there remains excess emission for both wavelength bands. The determination of the inclination angle of the accretion disk is an important discriminator between the different models explaining the narrow absorption lines. Assuming the excesses have a diskline profile, we can determine the inclination angle (Laor \cite{laor}), namely 80.9$^{\circ}$ $\pm$ 0.4, consistent with the 80.79$^{\circ}$ inclination angle measured for the host galaxy (Keel \cite{keel}). The emissivity slope is 1.9, similar to previously measured emissivity slopes. A Gaussian fit to these excesses is notably poorer, but from the UV band we know that broadened emission lines can be quite complex requiring several Gaussians to fit one line (Crenshaw et al. \cite{crenshaw03}). Further is the FW0I measured for the soft X-ray lines similar to the 13,000 km s$^{-1}$ as measured in the optical. It is thus unlikely that these excesses have diskline profiles.
\par
In the EPIC spectra we detect narrow Fe K$\alpha$ and \ion{Fe}{xxvi} Ly$\alpha$ emission lines. The intensities of these lines are consistent with the intensities measured with the earlier {\it Chandra} observation. The intensity of the Fe K$\alpha$ emission line is consistent between all observations since August 1997. This constant intensity and the narrowness of the line leads us to conclude that the line is probably formed at a large distance from the black hole.
Assuming that the Fe K$\alpha$ and \ion{Fe}{xxvi} emission lines have diskline profiles we find a very low inclination angle of 20$^{\circ}$ and a low emissivity slope of about 1, consistent with McKernan \& Yaqoob (\cite{mckernan}). The resulting profile is very similar to a Gaussian (dotted line in Fig.~\ref{fig:laor}). As a result we cannot distinguish from our spectra whether these lines have a diskline profile or not. As both determinations of the inclination angle discussed are inconsistent, we conclude that it is unlikely that disklines are detected in IC~4329A with current instrumentation.

\section{Summary}
IC~4329A has a heavily absorbed X-ray spectrum. In the complex high spectral resolution data studied, we detect seven distinct absorbers. From IC~4329A we detect absorption from the host galaxy, which is neutral; and absorption from the warm absorber closer to the nucleus. The warm absorber is similar to the warm absorber detected in NGC~5548, and has four different ionization components. The ionization parameters of the warm absorber components span at least two orders of magnitude. Similar to other Seyfert 1 galaxies, most of the gas in the warm absorber is highly ionized. We conclude that IC~4329A does not have a luke-warm absorber as suggested from a comparative study in the UV band. The lowest ionized component modeled here as absorption from the warm absorber could alternatively be also absorption from the host galaxy. If this is the case, then the gas in the host galaxy is lowly ionized instead of neutral. Two of the seven detected absorption components are not related to IC~4329A: the Galactic absorption and a moderately ionized absorber at redshift zero. 
\par
In our best fit to the data we need two broadened lines to fit the emission of the \ion{O}{viii} and \ion{N}{vii} Ly$\alpha$ lines. The fit with disklines is statistically better than a fit with two Gaussian lines. However, from broad emission line studies in the optical and UV band, we know that the line profile of these broad lines is complex, and poorly reproduced by a Gaussian line. Therefore, we conclude that the broadened lines are similar to the broad emission lines detected in the optical and UV band, and are not related to the disklines detected in some AGN. 
\par
We detect a narrow Fe K$\alpha$ line, and conclude from the lack of variability since the 1997 ASCA observation, that this line is probably formed at a distance of several pc from the nucleus. If we fit the Fe K$\alpha$ line with a diskline, we find parameters similar to those obtained from the HEG data (McKernan \& Yaqoob \cite{mckernan}). However, this profile can not be distinguished from a Gaussian line with current instrumentation, and we thus prefer the fit with a Gaussian line. 
\par
From the lightcurve we find that the flux varied by 17\%, and that the characteristic timescale for variability was between 120 ks and 190 ks. This is similar to previous measured characteristic timescales in this source. Consistent with earlier results, we did not detect variability in the softness ratio for this source, indicating that the soft and hard X-ray emission varied simultaneously.

\section*{ACKNOWLEDGMENTS}
 
The authors thank Alex Markovitz for his useful comments. This work is based on observations obtained with XMM-{\it Newton}, an ESA science mission with instruments and contributions directly funded by ESA Member States and the USA (NASA). SRON National Institute for Space Research is supported financially by NWO, the Netherlands Organization for Scientific Research. The MSSL authors acknowledge the support of the UK Particle Physics and Astronomy Research Council. E. Behar was supported by grant No. 2002111 from the United States-Israel Binational Science Foundation (BSF), Jerusalem, Israel.

\end{document}